\documentclass[12pt,a4paper,twoside]{article}

\usepackage{amsfonts,amsmath,amssymb,amsthm,latexsym}
\usepackage{times,mathptm}
\usepackage{handelingen}

\usepackage{ifpdf}
\ifpdf
\usepackage{epstopdf}
\usepackage{hyperref}
\else
\usepackage[hypertex]{hyperref}
\fi

\newtheorem{proposition}{Proposition}[section]

\newtheorem{theorem}[proposition]{Theorem}

\theoremstyle{definition}
\newtheorem{definition}[proposition]{Definition}

\theoremstyle{remark}

\newcommand{\thlabel}[1]{\label{th:#1}}
\newcommand{\thref}[1]{Theorem~\ref{th:#1}}
\newcommand{\selabel}[1]{\label{se:#1}}
\newcommand{\seref}[1]{Section~\ref{se:#1}}

\newcommand{\delabel}[1]{\label{de:#1}}
\newcommand{\deref}[1]{Definition~\ref{de:#1}}
\newcommand{\eqlabel}[1]{\label{eq:#1}}
\newcommand{\equref}[1]{(\ref{eq:#1})}

\newcommand\void[1]       {}

\newcommand\eps            {\varepsilon}
\newcommand\GV           {\mathcal{G\hspace{-2pt}V}}
\newcommand\Hom           {\mathrm{Hom}}
\newcommand\id            {{\rm id}}
\newcommand\nxt{\noindent\raisebox{.08em}{\rule{.44em}{.44em}}\hspace{.4em}}
\newcommand\one           {{\bf1}}
\newcommand\RS           {\mathcal{R\hspace{-2pt}S}}
\newcommand\RT           {\mathcal{R\hspace{-2pt}T}}
\newcommand\RTR           {\mathcal{R\hspace{-2pt}T}_{\hspace{-2pt}b}}
\newcommand\ST           {\mathcal{S\hspace{-1pt}T}}
\newcommand\STR         {\mathcal{S\hspace{-1pt}T}_{\hspace{-2pt}b}}
\newcommand\TV           {\mathcal{T\hspace{-3pt}V}}

\newcommand\Cb            {\mathbb{C}}

\newcommand\Rb            {\mathbb{R}}
\newcommand\Zb            {\mathbb{Z}}

\newcommand\Ac            {\mathcal{A}}
\newcommand\Bc            {\mathcal{B}}
\newcommand\Cc            {\mathcal{C}}
\newcommand\Hc            {\mathcal{H}}
\newcommand\Vc            {\mathcal{V}}

\newcommand\PF            {Z}

\renewcommand{\thefootnote}{\fnsymbol{footnote}}
\title{\MakeUppercase{Algebraic Structures in Euclidean and Minkowskian Two-Dimensional Conformal Field Theory}}
\author{Liang Kong$^{1}$ and Ingo Runkel$^{2}$}
\date{}
\renewcommand{\date}{\vspace{-5mm}}
\begin{document}
\maketitle \vspace*{-3mm}\relax
\renewcommand{\thefootnote}{\arabic{footnote}}

\noindent \textit{\small $^1$ California Institute of Technology,  
Center for the Physics of Information, Pasadena, CA 91125, USA\\
e-mail: lkong@caltech.edu }\\
\noindent \textit{\small $^2$ Department of Mathematics, King's College London 
  Strand, London WC2R 2LS, United Kingdom  \\
e-mail: ingo.runkel@kcl.ac.uk }

\begin{abstract}
\noindent 
We review how modular categories, and commutative and non-commutative Frobenius algebras arise in rational conformal field theory. For Euclidean CFT we use an approach based on sewing of surfaces, and in the Minkowskian case we describe CFT by a net of operator algebras. 
\end{abstract}

\vspace*{0.5em}

\section*{Introduction}
Since the seminal paper \cite{Belavin:1984vu}, two-dimensional conformal field theories (CFTs) have become a rich source of examples of solvable interacting quantum field theories. 

Physically, CFTs appear for example as scaling limits of two-dimensional statistical lattice models or one-dimensional quantum spin chains, and in the description of edge states in quantum Hall samples; CFTs also occur as world sheet theories in perturbative string theory. Mathematically, then, one would like to fix a set of axioms for CFTs and find interesting examples which can be proved to satisfy these axioms. 

\medskip

In \seref{def} we will briefly review two different definitions of CFT. The first will be referred to as `Euclidean CFT' and is based on the sewing of surfaces with parametrised boundaries. The surfaces are equipped with a conformal equivalence class of metrics of Euclidean signature, hence the name. The second approach will be called `Minkowskian CFT'  because it uses a net of operator algebras on two-dimensional Minkowski space. In \seref{chiral} we recall how in both approaches one can isolate a chiral subtheory, and how in favourable circumstances the representations of the chiral theories form a modular category. The problem of finding CFTs with prescribed chiral subtheories is then equivalent to locating appropriate algebra objects in the representation categories of the chiral theories. This holds for CFTs with and without boundaries as explained in \seref{closed} and \ref{se:open}. \seref{conc} contains our conclusions. 

As a word of caution, the authors' own background is on the side of Euclidean CFT, and while we have tried to resist our bias, the Euclidean part will still be more detailed than the Minkowskian one. Also, it was not our aim to give a historical account of the development together with a complete bibliography.

\newpage

\section{Definitions}\selabel{def}

\subsection{Euclidean CFT}\selabel{edef}

Axiomatic approaches to Euclidean CFT are based on formalising the properties of the operator product expansion or the behaviour of the string world sheet CFT under sewing \cite{Belavin:1984vu,Friedan:1986ua,Vafa:1987ea,Segal1988,Gawedzki1989,Gaberdiel:1998fs,HuKriz2004}. The definition of CFT we will start from is the one in \cite[\S\,4]{Segal1988}. Let $\RS$ be the category whose objects are ordered disjoint unions of circles $S^1$ and whose morphisms are equivalence classes of Riemann surfaces with a real analytic parametrisation of their boundaries. Two such surfaces are equivalent if there is a holomorphic isomorphism between them that is compatible with the boundary parametrisation. If we add permutations of the ordering of the circles in an object to the morphisms we obtain identity morphisms and a symmetric structure on $\RS$. Altogether $\RS$ is a symmetric monoidal category. Let $\TV$ be a symmetric monoidal category of locally convex and complete topological complex vector spaces. 

\begin{definition}\delabel{ecft}
An {\em oriented projective Euclidean CFT} is a continuous projective symmetric monoidal functor $\PF : \RS \rightarrow \TV$.
\end{definition}

Let $H = \PF(S^1)$. By a projective functor we mean a functor that assigns to a morphism $\Sigma : (S^1)^m \rightarrow (S^1)^n$ a complex line in $\Hom(H^{\otimes m}, H^{\otimes n})$, rather than a concrete element. This accounts for the conformal anomaly. There is an improved variant of \deref{ecft} which does not to use the word `projective' and where instead $\RS$ is equipped with a `rigging' and an appropriate power of the determinant line bundle. We would like to avoid going into this and refer to \cite[\S\,4\,\&\,App.\,B]{Segal1988} instead.

\deref{ecft} follows \cite[Def.\,4.2]{Segal1988}, except that we have allowed for an anomaly similar to \cite[Def.\,4.4]{Segal1988}, and that we do not demand the bilinear form and real structure on $\PF(S^1)$ or the compatibility of sewing with taking traces\footnote{
  The compatibility with taking traces does not follow directly from the axioms as it would 
  require that $\PF$ can be continuously extended to certain degenerate surfaces. 
  For example, $\PF$ applied to a cylinder of width zero (i.e.\ a circle)
  should be the identity map on $\PF(S^1)$. In any case, for \deref{gen0} below
  -- which is the one we will work with --
  compatibility with traces {\em does} follow from the axioms.}. 
Consequently, parts (iii)--(v) from \cite[Def.\,4.2]{Segal1988} are absent. 

\medskip

It turned out to be difficult to find examples of projective Euclidean CFTs. (As far as the authors can tell, apart from two-dimensional topological theories, none are known.)
Anticipating the following discussion, when starting the construction of projective Euclidean CFTs from vertex operator algebras, the three main problems are
\\[.3em]
(1)~We want a complete topological vector space $H = \PF(S^1)$. However the vertex operator algebra just gives us a countable direct sum $B = \bigoplus_{(h_l,h_r) \in \Rb^2} B_{h_l,h_r}$, where the grading is by eigenspaces of $L_0^{(l)}$ and $L_0^{(r)}$ (see \seref{ecft}). Denote by $\overline B$ the direct product $\prod_{(h_l,h_r) \in \Rb^2} B_{h_l,h_r}$, i.e.\ infinite sums are allowed. Then $B \subset H \subset \overline B$, and the question is which subset of $\overline B$ to choose and which topology to put on it.
\\[.3em]
(2)~Given a genus zero surface $\Sigma : (S^1)^m \rightarrow (S^1)^n$ we want a line of continuous linear maps $\PF(\Sigma): H^{\otimes m} \rightarrow H^{\otimes n}$. However, the approach via vertex operator algebras gives linear maps $B^{\otimes m} \rightarrow \overline{B^{\otimes n}}$. Thus when replacing $B$ and $\overline B$ by $H$ we have to make the source vector space bigger and the target vector space smaller in a consistent way.
\\[.3em]
(3)~We want $\PF(\Sigma \circ \Sigma') = \PF(\Sigma) \circ \PF(\Sigma')$. However, the composition is defined in terms of infinite sums over the graded components, and if $\Sigma \circ \Sigma'$ has genus $>0$ it is in general not known if these sums converge. 
\\[.3em]
For (1) and (2) a solution is known \cite{Huang2000} (it is not unique), but one can currently address (3)
only for certain vertex operator algebras and for compositions resulting in genus ${\le}1$ \cite{Zhu1996,Huang:2003cq}.

\medskip

The difficulty to find interesting examples of projective Euclidean CFTs motivates the introduction of a simpler object that can be studied as an intermediate step. In its definition we will avoid the above complications by restricting ourselves to genus zero surfaces, and by allowing categories with only partially defined composition. 

\medskip

A {\em sphere with tubes} is a Riemann sphere with two finite sets of marked points, called in-going and out-going punctures. No two marked points coincide, and each marked point $p$ has a local coordinate neighbourhood $(p,U,\varphi)$ where $p \in U$, $U$ is open, and $\varphi : U \rightarrow \Cb$ is an injective analytic map such that $\varphi(p)=0$. Given an out-going puncture $(p,U,\varphi)$ on a sphere $P$ and an in-going puncture $(q,V,\psi)$ on a sphere $Q$, we can define a glued sphere with tubes $Q {}_q\infty{}_p P$ as follows. The gluing is defined if there exists an $r>0$ such that $\varphi(U)$ contains the closed disc $\bar D_r$ of radius $r$, if $\psi(V)$ contains $\bar D_{1/r}$, and if $\varphi^{-1}(\bar D_r)$ and $\psi^{-1}(\bar D_{1/r})$ contain no further marked points. Then $Q {}_q\infty{}_p P$ is obtained by taking $P - \varphi^{-1}(D_r)$, with $D_r$ the open disc, and $Q - \psi^{-1}(D_{1/r})$, and identifying the points $\varphi(z) \sim \psi(-z^{-1})$ for all $|z|=r$, see \cite[Fig.\,2]{Huang-book} for an illustration. Two spheres with tubes are equivalent if there is a bijective analytic map between them that is compatible with the germs of the local coordinates. Different choices of $r$ in the gluing procedure lead to equivalent spheres with tubes.

The partial symmetric monoidal category $\ST$ (`spheres with tubes') has non-negative integers as objects, and the morphism set $\Hom(m,n)$ consists of equivalence classes of disjoint unions of spheres with tubes with a total of $m$ in-going punctures and $n$ out-going punctures, together with an ordering of the in- and out-going punctures (i.e.\ maps from $\{1,\dots,m\}$ to the set of in-going punctures and from $\{1,\dots,n\}$ to the set of out-going punctures). The composition is only partially defined, depending on the existence of appropriate $r$'s in the gluing procedure -- hence the qualifier `partial' above. In addition, we only allow a composition if the resulting surface is again a disjoint union of spheres; this excludes by hand the generation of higher genus surfaces. For example, we forbid the gluing of a sphere with two out-going punctures to a sphere with two in-going punctures.

The `annulus' $A_\lambda \in \Hom(1,1)$, for $\lambda \in \Cb^\times$, is defined to be $\Cb \cup \{\infty\}$ with in-going puncture $(0,\Cb,z \mapsto \lambda z)$ and out-going puncture $(\infty, \Cb^\times \cup \{\infty\}, z \mapsto \frac{-1}{z})$; it obeys $A_\lambda \circ A_\mu = A_{\lambda \cdot \mu}$. 

The partial symmetric monoidal category $\GV$ (`graded vector spaces') has as objects $\Rb\times\cdots\times\Rb$ graded vector spaces, $V = \bigoplus_{a} V_a$, where $a  \in \Rb^{n_V}$, $\dim(V_a) < \infty$, and the direct sum is countable. The morphisms $\Hom(V,W)$ are all linear maps from $V$ to $\overline W$, where $\overline W$ denotes the direct product instead of the direct sum. Again, the composition is only partially defined (and may not even be associative). The monoidal structure is such that $V \otimes W$ is $\Rb^{n_V + n_W}$-graded.

\begin{definition}\delabel{gen0}
An {\em oriented projective Euclidean genus-0 CFT} (genus-0 CFT for short) is a smooth projective symmetric monoidal functor $\PF : \ST \rightarrow \GV$ such that $B=\PF(1)$ is $\Rb \times \Rb$-graded, and $\PF(A_\lambda)$ is the line spanned by the linear map that acts on the graded component $B_{h_l,h_r} \subset B$ by multiplication with $\lambda^{-h_l} (\lambda^*)^{-h_r}$.
\end{definition}

The compatibility of the functor with composition is imposed only if the composition in $\ST$ is defined; in particular one demands that then the composition in $\GV$ is also defined.
By `smooth' we mean that the functor, restricted to any homogeneous subspace of the in-- and out-going vector spaces, depends smoothly on the moduli of the spheres with tubes.
The condition on $\PF(A_\lambda)$ requires in particular that $h_l-h_r \in \Zb$ whenever $B_{h_l,h_r} \neq \{0\}$, for otherwise $\lambda^{-h_l} (\lambda^*)^{-h_r}$ is not single-valued.

Again, one should really state \deref{gen0} without using the word `projective'. As for \deref{ecft}, this can be achieved by using appropriate powers of determinant line bundles over the moduli space of spheres with tubes. This is done in the original treatment in \cite[Sect.\,6.5]{Huang-book}, to which we refer for details (there the conditions are formulated in the language of operads). Furthermore, in \cite[Sect.\,2]{ko-ffa} the above geometric definition is linked to the purely algebraic notion of a `conformal full field algebra with nondegenerate invariant bilinear form', as defined in \cite[Def.\,1.19\,\&\,Sect.\,3]{ffa}.

One can prove that the usual suspects (Virasoro minimal models, WZW models, \dots) give examples of genus-0 CFTs, but it is also clear that \deref{gen0} is not quite general enough: it does not include logarithmic CFTs, such as the triplet model \cite{Gaberdiel:1998ps}, or non-compact CFTs, such as Liouville theory \cite{Teschner:2001rv}.

\subsection{Minkowskian CFT}\selabel{mcft}

We take as a starting point the axiomatic formulation of quantum field theory on Minkowski space in \cite{Haag:1963dh,Haag-book,Araki-book,Halvorson:2006wj} and describe its specialisation to conformal field theory on two-dimensional Minkowksi space $M_2$ following \cite{Brunetti:1992zf}. 

In light-cone coordinates $u = t + x$ and $v = t-x$ the Minkowski metric $dt^2-dx^2$ takes the form $d\hspace{-0.7pt}u \, d\hspace{-0.7pt}v$. The orientation and forward light-cone preserving conformal transformations of $M_2$ are given by two copies of the orientation preserving diffeomorphisms of the real line, $\mathrm{Diff}_+(\Rb) \times \mathrm{Diff}_+(\Rb)$, which act on the light cone coordinates as $(u,v) \mapsto (f(u),g(v))$. 

In quantum field theory a special role is played by symmetries which preserve the vacuum vector, and for most examples of CFTs the vaccum is not preserved by all of $\mathrm{Diff}_+(\Rb) \times \mathrm{Diff}_+(\Rb)$.
Instead we define the {\em conformal group} as the group generated by Poincar\'e transformations (translations, rotations, boosts), dilations $p \mapsto \lambda p$ ($\lambda>0$), and inversions $p \mapsto -p / p^2$ ($p^2 \neq 0$). This definition also applies in $d$-dimensional Minkowski space, for any $d$. The conformal group acts locally on $M_2$ and globally on the compactification $\bar M_2 = S^1 \times S^1$ (see \cite[Sect.\,1]{Brunetti:1992zf} for details). Define an action of $\mathrm{PSL}(2,\Rb)$ on $\Rb \cup \{\infty\}$ as
\begin{equation}
  g\,x = \frac{ax +b}{cx+d}
  \quad
  \text{where} \quad
  g = \begin{pmatrix} a&b \\ c&d \end{pmatrix} \in \mathrm{PSL}(2,\Rb) ~.
\end{equation}
The conformal group can be identified with $\mathrm{PSL}(2,\Rb) \times \mathrm{PSL}(2,\Rb)$, where an element $(g_1,g_2)$ maps a point $(u,v)$ in light cone coordinates to $(g_1 u, g_2 v)$.

A {\em double cone} $O$ is the open subset $O = \{ (x,t) \,|\, t+x \in I,\,t-x \in J \,\}$ of $M_2$, for $I,J \subset \Rb$ two open intervals. We will also use the notation $O = I \times J$. In $(x,t)$-coordinates a double cone looks like a diamond.

\begin{definition}\delabel{mcft}
A {\em Minkowskian CFT} $\Bc \equiv (\Hc,\Bc,\Omega,U)$ consists of a Hilbert space $\Hc$, a preferred vector $\Omega \in \Hc$ (the vacuum), a strongly continuous projective unitary positive energy representation $U$ of the conformal group on $\Hc$, and a map $\Bc : O \mapsto \Bc(O)$ (the local net of observables) which assigns to each double cone in $M_2$ a von Neumann algebra $\Bc(O)$ in $B(\Hc)$, the bounded linear operators on $\Hc$. 
This data satisfies:
\\[.3em]
(i) If $O_1 \subset O_2$ then $\Bc(O_1) \subset \Bc(O_2)$ (isotony).
\\[.3em]
(ii) If $O_1$ and $O_2$ are space-like separated (i.e.\ $p_1-p_2$ has length $<0$ for all $p_1 \in O_1$, $p_2 \in O_2$) then $[\Bc(O_1),\Bc(O_2)]= \{ 0 \}$ (causality).
\\[.3em]
(iii) If $O = \bigcup_n O_n$, where $O$ and the $O_n$ are double cones, then $\Bc(O) = \bigvee_n \Bc(O_n)$, where $\bigvee_n \Bc(O_n)$ is the von Neumann algebra generated by the $\Bc(O_n)$ (additivity).
\\[.3em]
(iv) $\Omega$ is invariant under the action of $\mathrm{PSL}(2,\Rb) \times \mathrm{PSL}(2,\Rb)$, and it is cyclic for $\bigcup_O \Bc(O)$.
\\[.3em]
(v) Let $O$ be a double cone and $g \in \mathrm{PSL}(2,\Rb) \times \mathrm{PSL}(2,\Rb)$ such that $gO$ exists (i.e.\ no point of $O$ gets mapped to infinity). Then
$U(g) \Bc(O) U(g)^* = \Bc(gO)$ (M\"obius covariance).
\end{definition}

The above definition is taken from \cite[Sect.\,1]{Brunetti:1992zf}, specialised to two dimensions. It does not include the full conformal covariance, which is different from the conformal group in one and two dimensions. The full conformal covariance does not play a role for the algebraic properties we are going to describe, but it can be included (and indeed should be for comparison with \deref{gen0}, as it is the analogue of an annulus with arbitrary local coordinates). One would then add the following to \deref{mcft} (adapted from \cite[Sect.\,2]{Kawahigashi:2002px}): There is a projective unitary representation $\tilde U$ of $\mathrm{Diff}_+(S^1) \times \mathrm{Diff}_+(S^1)$ on $\Hc$ extending that of $\mathrm{PSL}(2,\Rb) \times \mathrm{PSL}(2,\Rb)$ (via the common action on the compactification $\bar M_2 = S^1 \times S^1$). Property (v) now holds for $g \in \mathrm{Diff}_+(S^1) \times \mathrm{Diff}_+(S^1)$, and in addition one demands that if $g \in \mathrm{Diff}_+(S^1) \times \mathrm{Diff}_+(S^1)$ leaves a double cone $O$ point-wise fixed, then $\tilde U(g) A \tilde U(g)^* = A$ for all $A \in \Bc(O)$. 

\medskip

The formulation of CFT via nets of operator algebras has the nice feature that it is wywiwyg (what-you-want-is-what-you-get), and an analogous simplification as that from \deref{ecft} to \ref{de:gen0} is not necessary. One way to understand this is to note that in \deref{mcft} one never deals with higher genus surfaces, and in this sense \deref{mcft} corresponds to \deref{gen0}, rather than to \deref{ecft}.

The basic idea how Definitions \ref{de:gen0} and \ref{de:mcft} are related is as follows, cf.\ \cite[Ch.\,II.2]{Haag-book}. One supposes that the net of von Neumann algebras is obtained from fields satisfying the Wightman axioms. The vacuum correlators of these fields are called Wightman functions and allow for analytic continuation to imaginary time. These analytically continued functions define the correlators of a Euclidean theory. For the converse direction one has to require that the Euclidean correlators obey Osterwalder-Schrader reflection positivity (this is not imposed in \deref{gen0} as it would in particular exclude non-unitary CFTs).

However, the authors are not aware of a theorem which makes a precise statement about the relation between Definitions \ref{de:gen0} and \ref{de:mcft}.

\medskip

As a final comment, let us note that one can replace the formulation of Minkowskian quantum field theory in terms of nets of operator algebras by a formulation in terms of functors \cite{Schreiber:2008uk}. The two Definitions \ref{de:ecft} and \ref{de:mcft} then use the same language.

\section{Chiral conformal field theory}\selabel{chiral}

Here we will explain how to isolate chiral subtheories of a CFT and recall that in favourable circumstances -- loosely called `rational' -- the representations of such a chiral CFT form a modular category. To begin with, let us state the definition of the latter.

\begin{definition} \cite{brug2}
A category $\Cc$ is called {\em premodular} with ground field $\Cb$ if it is an 
abelian $\Cb$-linear semisimple category with finitely many isomorphism classes 
of simple objects, endowed with a ribbon structure, and with simple monoidal unit.
\end{definition}

The ribbon structure on $\Cc$ consists of a tensor product, a braiding, a twist, and compatible dualities, see e.g.\ \cite[Sect.\,I.1]{Turaev-book}. The braiding is denoted by $c_{U,V} : U \otimes V \rightarrow V \otimes U$.

\begin{definition}\delabel{MTC}
A {\em modular category} is a premodular category with the following property: if for a fixed simple object $U$ one has $c_{U,V} = (c_{V,U})^{-1}$ for all simple objects $V$, then $U \cong \one$. 
\end{definition}

This is related to the original definition in \cite[Sect.\,II.1.4]{Turaev-book} via \cite[Prop.\,1.3]{Beliakova2000}. A modular category provides precisely the data needed to define a three-dimensional topological field theory \cite{retu,Turaev-book}. The relation to chiral CFT in fact precedes this definition and originated in the study of three-dimensional Chern-Simons theories \cite{Witten:1988hf,Frohlich:1989gr} and the factorisation and monodromy properties of conformal blocks \cite{Moore:1988qv}. 

For Minkowskian CFTs we will also need *-operations (see e.g.\ \cite[Sect.\,2.4]{Muger2001aj}) and unitary modular categories.

\begin{definition}\delabel{star}
A $*$-{\em category} is a $\Cb$-linear category equipped with a family of antilinear involutions $(\,\cdot\,)^* :  \Hom(U,V) \rightarrow \Hom(V,U)$ which are positive in the sense that for any morphism $s : U \rightarrow V$, $s^* \circ s = 0$ implies $s=0$. A {\em monoidal} $*$-{\em category} is a monoidal category which is a *-category such that for all morphisms $s : U \rightarrow V$ and $t : U' \rightarrow V'$ one has $(s \otimes t)^* = s^* \otimes t^*$.
\end{definition}

A {\em unitary modular category} is a modular category which is  a monoidal *-category, and where the $*$-operation satisfies a set of compatibility relations with braiding, twist, and the duality morphisms (see \cite[Sect.\,2.4]{Muger2001aj} and \cite[Sect.\,II.5]{Turaev-book} for details).

\subsection{Euclidean CFT}\selabel{ecft}

Recall that the Virasoro algebra $\mathrm{Vir}$ is the complex Lie algebra with basis elements $C$ and $L_m$, $m \in \Zb$, where $C$ is central and the $L_m$ have Lie brackets
\begin{equation}
  [L_m, L_n] = (m-n) \, L_{m+n} + \tfrac{1}{12} (m^3-m) \, \delta_{m+n,0} \, C ~.
\end{equation}
The elements $\{L_{-1}, L_0, L_1 \}$ span a subalgebra isomorphic to $sl(2,\Cb)$. Let $\PF$ be a genus-0 CFT. By evaluating $\PF$ on two-punctured spheres in $\Hom(1,1)$ with different local coordinates one obtains an action of two copies of the Virasoro algebra\footnote{
  To be precise, this result needs that $\PF$ comes from a non-projective functor using appropriate powers of the determinant line bundle, which then fixes the central charges $c^l$ and $c^r$.},
 $\mathrm{Vir}^{(l)} \oplus \mathrm{Vir}^{(r)}$, on $B = \PF(1)$ (this follows from \cite[Prop.\,5.4.4]{Huang-book} and \cite[Thm.\,1.19]{ko-ffa}). The central elements $C^{(l)}$ and $C^{(r)}$ are represented by constants $c^l, c^r \in \Cb$ (the left and right central charge), and $L_0^{(l)}$ and $L_0^{(r)}$ act as $h_l \cdot \id$ and $h_r \cdot \id$ on the homogeneous subspace $B_{h_l,h_r}$, respectively.
We call an element of $B$ {\em chiral} if it is annihilated by either one of the two copies of $sl(2,\Cb)$ in $\mathrm{Vir}^{(l)} \oplus \mathrm{Vir}^{(r)}$\footnote{
 In fact it is enough to demand this for $L_{-1}^{(l)}$ or $L_{-1}^{(r)}$, cf.\ \cite[Cor.\,4.7.6]{Lepowski-Li-book}.}. 

By a {\em subtheory} of $\PF$ we mean a graded subspace $\tilde B \subset B$ such that $\PF$ restricts to a genus-0 CFT $\tilde\PF$ with $\tilde\PF(1) = \tilde B$. In other words, the inclusion $\tilde B \subset B$ provides a natural monoidal transformation $\tilde\PF \rightarrow \PF$. A subtheory where $sl(2,\Cb)^{(r)}$ acts trivially does depend holomorphically on the moduli of $\ST$, while a subtheory where $sl(2,\Cb)^{(l)}$ acts trivially depends anti-holomorphically on the moduli. We call such subtheories {\em chiral}. 

\medskip

The algebraic structure which encodes a chiral genus-0 CFT is called a {\em vertex operator algebra}. It consists of a tuple $(V,Y,\one,T)$ where $V = \bigoplus_{n \in \Zb} V_{(n)}$ is a $\Zb$-graded complex vector space such that $\dim V_{(n)} < \infty$ and $V_{(n)} = 0$ for $n \ll 0$. $Y$ is a linear map $Y(\,\cdot\,,z) : V \rightarrow \mathrm{End}(V)[\![z^{\pm 1}]\!]$ from $V$ to formal power series in $z$ and $z^{-1}$ with coefficients in $\mathrm{End}(V)$. Finally, $\one \in V_{(0)}$ (the vacuum) and $T \in V_{(2)}$ (the stress tensor) are preferred elements of $V$. These data have to satisfy a list of conditions, for which we refer to \cite[Def.\,3.1.1\,\&\,3.1.22]{Lepowski-Li-book}.

A {\em module} over a vertex operator algebra $V$ is a pair $(M,Y_M)$, where $M$ is a $\Cb$-graded vector space and $Y_M(\,\cdot\,,z) : V \rightarrow \mathrm{End}(M)[\![z^{\pm 1}]\!]$ is a linear map, satisfying the conditions listed in \cite[Def.\,4.1.1\,\&\,4.1.6]{Lepowski-Li-book}. $V$ is a module over itself, and for every module $M$ one can define the {\em contragredient module} $M^\vee$ on the graded dual of $M$, see \cite[Sect.\,5.2]{FHL-book}.

\begin{theorem}\thlabel{VOA-CFT}
A vertex operator algebra $V$ with $V^\vee \cong V$ as $V$-modules defines a genus-0 CFT $\PF_{V,\one}$. It depends holomorphically on the moduli of $\ST$ and obeys $\PF_{V,\one}(1) = V$, where the subspace $V_{(m)} \subset V$ has $\Rb \times \Rb$ grade $(m,0)$.
\end{theorem}

This theorem is implied by \cite[Prop.\,5.4.1]{Huang-book} and \cite[Thm.\,2.7]{ko-ffa}. In fact, these references make a stronger statement: If one refines the conditions of the theorem (in particular one has to work with the non-projective variant of the genus-0 CFT) one obtains an equivalence of categories between appropriate vertex operator algebras and appropriate Euclidean genus-0 CFTs. For example, the vector $\one \in V_{(0)}$ is given by applying the functor $Z$ to a sphere $\Cb \cup \{\infty\}$ with one out-going puncture at $\infty$ and standard coordinate $z \mapsto -1/z$, the vector $T \in V_{(2)}$ is obtained from the same sphere but with a deformed coordinate, and the sphere with two in-going punctures at $0$ and $z$ and one out-going puncture at $\infty$ gives rise to $Y(\,\cdot\,,z)$.

A vertex operator algebra $V$ as in \thref{VOA-CFT} also defines a genus-0 CFT $Z_{\one,V}$ which depends {\em anti}-holomorphically on the moduli of $\ST$ by taking complex conjugates in the appropriate places. 

\medskip

The category of modules over a (sufficiently nice) vertex operator algebra, $\mathrm{Rep}(V)$, carries the structure of a braided tensor category \cite{hl-vtc}. The following theorem makes the connection to \deref{MTC}, and we refer to \cite{Huang2005} for an explanation of the technical terms appearing in its formulation.

\begin{theorem}\thlabel{voa-mc} \cite[Thm.\,4.6]{Huang2005}
Let $V$ be a simple vertex operator algebra satisfying the following conditions:
(i) $V_{(n)} = 0$ for $n<0$, $V(0) = \Cb \one$ and $V^\vee  \cong V$ as $V$-module,
(ii) every $\mathbb{N}$-gradable weak $V$-modules is completely reducible, and
(iii) $V$ is $C_2$-cofinite. Then $\mathrm{Rep}(V)$ is a modular category.
\end{theorem}

For the sake of this paper, we will refer to vertex operator algebras which satisfy the conditions in the preceding theorem as {\em rational}.

\subsection{Minkowskian CFT}

Let $(\Hc,\Bc,\Omega,U)$ be a Minkowskian CFT and recall the notation $O = I \times J$ for double cones from \seref{mcft}. By a chiral observable on $O$ we mean an element of $\Bc(O)$ which commutes with the action of one of the two copies of $PSL(2,\Rb)$ in the conformal group. Let us concentrate on chiral observables which commute with the right copy and denote the set of all such chiral observables by $\Vc_L(O)$
\cite[Def.\,2.1]{Rehren:1999ab}:
\begin{equation}
  \Vc_L(I \times J) = \big\{ \, A \in \Bc(I \times J) \,\big|\, [A , U(\id \times g) ] = 0 ~\text{for~all}~ g \in PSL(2,\Rb) \big\} ~.
\end{equation}  
Since for fixed $I$, any two double cones $I \times J_1$ and $I \times J_2$ are related by a M\"obius transformation of the form $\id \times g \in \mathrm{PSL}(2,\Rb) \times \mathrm{PSL}(2,\Rb)$, by definition of $\Vc_L$ and by M\"obius covariance of the net $\Vc$ we have $\Vc_L(I \times J_1) = \Vc_L(I \times J_2)$. It is therefore consistent to define $\Vc_L(I) \equiv  \Vc_L(I \times J)$, and we obtain a net $\Vc_L$ on $\Rb$ (identified with the light ray $\{x=t\}$). The net $\Vc_L$ satisfies $[\Vc_L(I_1),\Vc_L(I_2)]=\{0\}$ for all disjoint open intervals $I_1$ and $I_2$, and it is M\"obius covariant, cf.\,\cite[Sect.\,2]{Rehren:1999ab}. 

A subtheory of $(\Hc,\Bc,\Omega,U)$ is a Minkowskian CFT $(\tilde\Hc,\tilde\Bc,\tilde\Omega,\tilde U)$ such that $\tilde\Hc \subset \Hc$, $\tilde\Bc(O) \subset \Bc(O)$, $\tilde\Omega = \Omega$, and $\tilde U = U|_{\tilde\Hc}$. For example, $\Vc_L$ is a subtheory of  $(\Hc,\Bc,\Omega,U)$ (the Hilbert space can be obtained from \cite[Lem.\,2.2]{Rehren:1999ab}).

Because of M\"obius covariance it is convenient to compactify the light ray $\{ x=t \}$ to $S^1$, motivating the following definition to capture the properties of the chiral observables.

A {\em M\"obius covariant net on} $S^1$ is a tuple $\Vc \equiv (\Hc,\Vc,\Omega,U)$, where $\Hc$ is a separable Hilbert space, $\Vc : I \mapsto \Vc(I)$ a map from open, non-dense intervals in $S^1$ to von Neumann algebras on $\Hc$, $U$ is a strongly continuous (non-projective) unitary positive energy representation of $PSL(2,\Rb)$ on $\Hc$, and $\Omega \in \Hc$ is an invariant vector with respect to that action.
The data $\Hc,\Vc,U,\Omega$ has to satisfy a number of conditions similar to \deref{mcft}, for details 
we refer to \cite[Def.\,2.3\,\&\,2.5]{Gabbiani:1992ar} or \cite[Sect.\,2.1]{Kawahigashi:2002px}.

The net $\Vc$ is called {\em irreducible} if $\bigvee_{I} \Vc(I) = B(\Hc)$, i.e.\ if the von Neumann algebra generated by all $\Vc(I)$ is equal to the bounded linear operators on $\Hc$. 

A {\em representation} of a M\"obius covariant net $\Vc$ is a tuple $(\Hc_\pi,\pi,U_\pi)$, where $\Hc_\pi$ is a separable Hilbert space, $\pi$ is a map which assigns to every open non-dense interval in $S^1$ a representation $\pi_I$ of the algebra $\Vc(I)$ on $\Hc_\pi$, and $U_\pi$ is a strongly continuous projective unitary positive energy representation of $PSL(2,\Rb)$ on $\Hc_\pi$. Again, these data are subject to conditions for which we refer to \cite[Def.\,4.1]{Gabbiani:1992ar}. We denote the category of representations of $\Vc$ by $\mathrm{Rep}(\Vc)$. It is a braided monoidal *-category (see e.g.\ \cite[Sect.\,8]{Halvorson:2006wj}).

\begin{theorem}\thlabel{klm} \cite{Kawahigashi:1999jz}
Let $\Vc$ be an irreducible M\"obius covariant net on $S^1$ which is strongly additive, split, and such that the inclusion of $\Vc(E)$ into $\Vc(E')'$ has finite index for each non-dense $E \subset S^1$ which is the union of two open intervals. Then $\mathrm{Rep}(\Vc)$ is a unitary modular category.
\end{theorem}

Here $E'$ denotes the open set $S^1 - \overline E$, $A'$ is the commutant of a von Neumann algebra $A$, and we refer to \cite{Kawahigashi:1999jz} for the explanation of the terms `strongly additive', `split' and of the index. A net with the extra properties listed in \thref{klm} is called completely rational \cite[Def.\,8]{Kawahigashi:1999jz}. It is proved in \cite{Kawahigashi:1999jz} that $\mathrm{Rep}(\Vc)$ is semisimple with a finite number of simple objects (Cor.\,10\,\&\,39), that the simple objects have finite dimension (Cor.\,10) so that $\mathrm{Rep}(\Vc)$ has duals \cite[Sect.\,8.3]{Halvorson:2006wj}, and that the braiding is non-degenerate (Cor.\,37). The property `modular PCT' needed in \cite{Kawahigashi:1999jz} is automatic under the assumptions in \thref{klm}, cf.\,\cite[Sect.\,3]{Kawahigashi:1999jz}).

\section{Closed conformal field theory}\selabel{closed}

A `bottom up' approach to constructing a CFT is to first decide on a chiral theory, and then to study which CFTs can be build on top of it. It turns out that this question can be studied in the category of representations of the chiral theory. To this end we need the notion of a Frobenius algebra in a monoidal category.

\begin{definition}\delabel{frob}
A {\em Frobenius algebra} in a monoidal category $\Cc$ is a tuple $A \equiv (A,m,\eta,\Delta,\eps)$ where $A \in \Cc$, 
$m \in \Hom(A \otimes A, A)$,
$\eta \in \Hom(\one, A)$,
$\Delta \in \Hom(A, A \otimes A)$ and
$\eps \in \Hom(A, \one)$, subject to the relations
\\[.3em]
$m \circ (\id_A \otimes \eta) = \id_A =  m \circ (\eta \otimes \id_A)$~~(unit), \\
$(\id_A \otimes \eps) \circ \Delta = \id_A = (\eps \otimes \id_A) \circ \Delta$~~(counit),  \\
$m \circ (m \otimes \id_A) = m \circ (\id_A \otimes m)$~~(associativity),  \\
$(\Delta \otimes \id_A) \circ  \Delta = (\id_A \otimes \Delta) \circ \Delta$~~(coassociativity),  \\
$(m \otimes \id_A) \circ (\id_A \otimes \Delta) = \Delta \circ m = (\id_A \otimes m) \circ (\Delta \otimes \id_A)$~~(Frobenius).
\end{definition}

The notions to be introduced below need more structure than just a tensor product on the underlying category, but not the full data of a modular category (or that of a Frobenius algebra, for that matter). As we will only use them in the latter setting, we prefer not state each definition in its minimal environment. We will denote the duality morphisms in a modular category by
$b_U : \one \rightarrow U \otimes U^\vee$,
$d_U : U^\vee \otimes U \rightarrow \one$,
$\tilde b_U : \one \rightarrow U^\vee \otimes U$ and
$\tilde d_U : U \otimes U^\vee \rightarrow \one$.
The following definitions can be found in \cite[Sect.\,2\,\&\,4]{Fuchs:2001qc} and \cite[Sect.\,2]{tft0}

\begin{definition}
Let $\Cc$ be a modular category. A Frobenius algebra $A$ in $\Cc$ is called\\
- {\em symmetric} if the following two morphisms from $A$ to $A^\vee$ are equal
$$
  \Phi_1 = ((\eps\circ m)\otimes \id_{A^\vee}) \circ (\id_A\otimes b_A) ~~,~~~
  \Phi_2 = (\id_{A^\vee}\otimes(\eps\circ m)) \circ (\tilde b_A\otimes \id_A) ~,
$$
- {\em special} if $m \circ \Delta = \beta_A \id_A$ and $\eps \circ \eta = \beta_\one \id_\one$ for some non-zero $\beta_A$, $\beta_\one \in \Cb$,
\\
- {\em commutative} if $m \circ c_{A,A} = m$,
\\
- {\em simple} if it is simple as an $A$-$A$-bimodule,
\\
- {\em haploid} if $\Hom(\one,A) = \Cb \, \eta$.
\end{definition}

The Frobenius algebras for closed CFTs will live in products of the categories of representations obtained via the construction in \seref{chiral}. Given two modular categories $\Cc$ and $\mathcal{D}$ we write $\Cc\boxtimes\mathcal{D}$ for the modular category whose objects are direct sums of pairs of objects and whose morphism spaces are tensor products over $\Cb$ of those of $\Cc$ and $\mathcal{D}$. We denote by $\Cc_-$ the modular category obtained from $\Cc$ by replacing the braiding and twist by their inverses. For symmetry of notation we also set $\Cc_+ \equiv \Cc$.

\medskip

When discussing genus 1 conditions for the Euclidean CFT we will need the following -- at first sight not very illuminating -- notion (this is essentially \cite[Def.\,6.6]{Kong2006c}; we give the formulation in \cite[Def.\,3.1]{Kong:2008ci}, where one can also find a pictorial representation).

\begin{definition}\delabel{modinv}
An algebra $B$ in a modular category $\Cc$ is {\em modular invariant} if $\theta_B = \id_B$ and for all $W \in \Cc$ the following equality of morphisms $B \otimes W \rightarrow W$ holds: 
$$
\begin{array}{l}\displaystyle
  \big[ \tilde d_B \otimes \id_W \big] 
  \circ \big[ m \otimes (c_{W,B^\vee} \circ c_{B^\vee,W}) \big]
  \circ \big[ \id_B \otimes b_B \otimes \id_W \big]
\\[.5em] \displaystyle
  = 
  \sum_{k} \frac{\dim(U_k)}{\mathrm{Dim}(\Cc)^{\frac12}}
  \big[ \tilde d_B \otimes \tilde d_{U_k} \otimes \id_W \big] 
  \circ \big[ m \otimes (c_{U_k,B^\vee} \circ c_{B^\vee,U_k}) \otimes (c_{W,U_k^\vee} \circ c_{U_k^\vee,W}) \big]
\\[-.3em] \displaystyle
  \hspace{25em} \circ \big[ \id_B \otimes b_B \otimes b_{U_k} \otimes \id_W \big]~~.
\end{array}
$$
\end{definition}

Here the sum runs over the labels of a set of representatives $\{ U_k \}$ of the isomorphism classes of simple objects in $\Cc$, and $\mathrm{Dim}(\Cc)$ $=$ $\sum_k \dim(U_k)^2$. For a (pre)modular category $\Cc$ one has $\mathrm{Dim}(\Cc) \ge 1$ \cite[Thm.\,2.3]{eno}.
$\theta_B : B \rightarrow B$ denotes the twist and is part of the data of the modular category.
A commutative symmetric Frobenius algebra $B$ also obeys $\theta_B = \id_B$ \cite[Prop.\,2.25]{Frohlich:2003hm}. 

For a haploid commutative special symmetric Frobenius algebra, modular invariance can be replaced by a much simpler condition on the quantum dimension.

\begin{theorem} \thlabel{mod-dim} 
\cite[Thm.\,3.4]{Kong:2008ci}\footnote{
  In \cite{Kong:2008ci} both, \deref{modinv} and \thref{mod-dim}, are formulated in the special case that $\Cc = \mathcal{D}_+ \boxtimes \mathcal{D}_-$ for a modular category $\mathcal{D}$, but this condition is not needed for the definition, nor used in the proof.}
Let $B$ be a haploid commutative symmetric Frobenius algebra in a modular category $\Cc$. Then $B$ is modular invariant if and only if $\dim(B) = \mathrm{Dim}(\Cc)^{\frac12}$.
\end{theorem}

From \cite[Thm.\,4.5]{Kirillov:2001ti} one concludes that a haploid commutative special symmetric Frobenius algebra $B \in \Cc$ obeys $\dim(B) \le \mathrm{Dim}(\Cc)^{\frac12}$. Thus $B$ is modular invariant if and only if it is of maximal dimension.

\medskip

For Minkowskian CFT we will require compatibility of the data defining a Frobenius algebra with the *-operation.

\begin{definition}\delabel{*-frob}
A Frobenius algebra $(A,m,\eta,\Delta,\eps)$ in a monoidal *-category for which $m^* = \Delta$ and $\eta^* = \eps$
is called a $*$-{\em Frobenius algebra}.
\end{definition}

\subsection{Euclidean CFT}\selabel{ecft-c-0}

Given a vertex operator algebra $V$ satisfying the conditions in \thref{VOA-CFT} we have obtained a genus-0 CFT with $Z_{V,\one}(1) = V$, where $V_{(m)}$ has $\Rb \times \Rb$-grade $(m,0)$, and which depends holomorphically on the moduli of $\ST$. This can be combined with the corresponding anti-holomorphic theory to give the following theorem (\cite[Prop.\,1.6]{ffa} and \cite[Thm.\,2.7]{ko-ffa}).

\begin{theorem}
Two  vertex operator algebras $V_L$ and $V_R$ such that $V_L^\vee \cong V_L$ and $V_R^\vee \cong V_R$ as modules give rise to a genus-0 CFT $Z_{V_L,V_R}$ with $Z_{V_L,V_R}(1) = V_L \otimes V_R$, and where the subspace $(V_L)_{(m)} \otimes (V_R)_{(n)} \subset V_L \otimes V_R$ has $\Rb \times \Rb$ grade $(m,n)$.
\end{theorem}

We can now try to classify theories which have $Z_{V_L,V_R}$ as a subtheory. This has been carried out in the case that both $V_L$ and $V_R$ are rational in the sense of \thref{voa-mc}. With the ingredients we have introduced, we can state one of the two directions as a theorem.

\begin{theorem}\thlabel{csFa-closed-0}
Let $V_L$ and $V_R$ be rational vertex operator algebras. A commutative symmetric Frobenius algebra $B$ in $\mathrm{Rep}(V_L)_+ \boxtimes \mathrm{Rep}(V_R)_-$ gives rise to a genus-0 CFT which has $Z_{V_L,V_R}$ as a subtheory. 
\end{theorem}

This follows from \cite[Thm.\,4.15]{ko-ffa}, were, for the non-projective version of a genus-0 CFT, an equivalence between such 	CFTs and commutative symmetric Frobenius algebras in $\mathrm{Rep}(V_L)_+ \boxtimes \mathrm{Rep}(V_R)_-$ is proved. 

\subsubsection*{Genus 1}

As an intermediate step on the way from \deref{gen0} to \deref{ecft} one can modify the source category $\ST$ of the functor $\PF : \ST \rightarrow \GV$ to the category of `Riemann surfaces with tubes', $\RT$, which is defined in the same way as $\ST$ in \seref{edef}, but with the underlying surface in a morphism not restricted to genus 0, and with composition not restricted to result in genus 0 surfaces (this has been called `partial CFT' in  \cite[sect.\,1]{Kong2006c}). In this way one avoids the topological completion of $B = \PF(1)$, but one needs to address the convergence of traces. Again, this is currently too difficult, but some progress can be made at genus 1.

Every higher genus Riemann surface can be obtained by sewing spheres with punctures, and so if a genus-0 CFT $\PF : \ST \rightarrow \GV$ can be extended to $\tilde\PF : \RT \rightarrow \GV$, this extension is {\em unique}. However, the existence of $\tilde\PF$ will impose extra conditions on $\PF$. 

A necessary condition can be obtained as follows \cite{Cardy:1986ie}. Take a sphere $S_{0,z}^\infty \in \Hom_{\ST}(2,1)$ with two in-going and one out-going puncture. Denote by $s(S_{0,z}^\infty) \in \Hom_{\RT}(1,1)$ the surface obtained by sewing the punctures at $0$ and $\infty$. Then the axioms require $\tilde\PF( s(S_{0,z}^\infty) ) = \mathrm{tr}_B \PF(S_{0,z}^\infty)$. One has to ensure that the traces $\mathrm{tr}_B \PF(S_{0,z}^\infty)$ converge, and that $\tilde\PF( s(S_{0,z}^\infty) )$ only depends on the conformal equivalence class of $s(S_{0,z}^\infty)$, leading to a modular invariance condition. Under the assumptions of \thref{csFa-closed-0} and for particular choices of local coordinates around $0,z,\infty$, in \cite{Huang:2006ar} and \cite[Thm.\,6.7]{Kong2006c} this was shown to be equivalent to the condition of modular invariance on the commutative symmetric Frobenius algebra $B$ as given in \deref{modinv}.

Let us conclude with some remarks.
\\[.3em]
\nxt The modular invariance condition in \deref{modinv} is a necessary condition for $\PF : \ST \rightarrow \GV$ to extend to $\tilde\PF : \RT \rightarrow \GV$. While it is not known if it is sufficient, by the reasoning in \cite{Sonoda:1988} and by the relation to three-dimensional topological field theory (see \seref{ecft-oc-0}) it should be sufficient {\em provided} the conditions in \thref{voa-mc} already ensure the convergence of the higher genus compositions.
\\[.3em]
\nxt Someone not interested in string theory may wonder why one should bother with higher genus conditions. Here is a reason: An important source of Euclidean CFTs are continuum limits of lattice models at a critical point. While it might be difficult to build in the laboratory, these models are typically well-defined on lattices with periodic boundary conditions and modular invariant before taking the continuum limit (an example is the Ising model on a square lattice). So one would expect the continuum theory to preserve this property, and in the previous point we have seen that there should be no further conditions beyond genus 1.
\\[.3em]
\nxt For two-dimensional {\em topological} field theories one can construct functors as in \deref{ecft}, and they are in one-to-one correspondence with commutative Frobenius algebras over the complex numbers, see e.g.\ \cite{Kock-book}. The modular invariance condition in \deref{modinv} is trivially true in this case and is only discovered once non-trivial dependence on the complex structure moduli is included.

\subsection{Minkowskian CFT}\selabel{mcft-c}

By considering observables which commute with one of the two copies of $PSL(2,\Rb)$ we were led to the notion of a M\"obius covariant net on $S^1$, the compactified light ray. Taking two such nets $\Vc_L$ and $\Vc_R$, one for each light ray through zero, we obtain a Minkowskian CFT $\Vc_L \otimes \Vc_R$ as in \deref{mcft} which assigns to a double cone $I \times J$ the algebra $\Vc_L(I) \otimes \Vc_R(J)$ of bounded operators on $\Hc_L \otimes \Hc_R$. We can now ask which Minkowskian CFTs contain $\Vc_L \otimes \Vc_R$ as a subtheory.

\begin{theorem}\thlabel{m-c-frob}
Let $\Vc_L$ and $\Vc_R$ be two completely rational nets on $S^1$. There is a one-to-one correspondence between haploid commutative symmetric *-Frobenius algebras $B$ in $\mathrm{Rep}(\Vc_L)_+ \boxtimes \mathrm{Rep}(\Vc_R)_-$ and Minkowskian CFTs $\Bc$ containing $\Vc_L \otimes \Vc_R$ as a subtheory of finite index.
\end{theorem}

The theorem follows from \cite[Thm.\,4.9]{Longo:1994xe} and \cite[Prop.\,3.2]{Rehren:1999ab}. While the above formulation is the one we are interested in here, the results in \cite{Longo:1994xe,Rehren:1999ab} hold under much weaker assumptions, and we refer to these papers for details. Also, in  \cite{Longo:1994xe,Rehren:1999ab} the notion of Q-systems is used to characterise finite-index inclusions of subfactors. The relation to Frobenius algebras is described e.g.\ in \cite[Sect.\,6.4]{Muger2001aj}.

Comparing Theorems \ref{th:csFa-closed-0} and \ref{th:m-c-frob} it seems that the Minkowskian CFT needs stronger conditions on the Frobenius algebra. This is indeed true, but it is directly related to the physical interpretation. The compatibility with the *-operation is ultimately linked to the formulation of the theory in terms of operators on a Hilbert space (read: the authors do not know a clear and short reason). Haploidity is easier to interpret: it amounts to uniqueness (up to multiples) of the vacuum vector. 

\subsubsection*{Genus 1}

Modular invariance is at first sight not a natural condition to expect (or impose) in the Min\-kowskian setting. Physically, it is the statement\footnote{
  More precisely, this amounts to $S$-invariance. $T$ invariance follows from locality, i.e.\ the fact that the Frobenius algebra in \thref{m-c-frob} is symmetric and commutative.} 
that (in appropriate units) the thermal partition function at inverse temperature $\beta$ of the quantum system placed on a circle of radius $R$ is equal to that of the same system on a circle of radius $\beta$ at inverse temperature $R$.

However, somewhat surprisingly, from the point of view of {\em boundary} Minkowskian CFT there is a natural condition -- Haag duality -- which implies modular invariance. We will get back to this in \seref{mcft-oc}.

\section{Open/closed conformal field theory}\selabel{open}

The algebraic structure we introduce here, namely that of a Cardy-algebra, is tailored more to the Euclidean approach. The relation with Minkowskian CFT comes about when adding natural assumptions such as uniqueness of the vacuum.

\medskip

Let $\Cc$ be a modular category. The tensor product provides a functor $T : \Cc_+ \boxtimes \Cc_- \rightarrow \Cc$ and the braiding on $\Cc$ allows to turn $T$ into a monoidal functor (see \cite[Sec.\,2.4]{Kong:2008ci} for details). The functor $T$ has a two-sided adjoint $R : \Cc \rightarrow \Cc_+ \boxtimes \Cc_-$, which acts on objects $V \in \Cc$ and morphisms $f : V \rightarrow W$ as 
\begin{equation}
  R(V) = \bigoplus_{k} (V \otimes U_k^\vee) \times U_k 
  ~~,~~~
  R(f) = \bigoplus_{k} (f \otimes \id_{U_k^\vee}) \times \id_{U_k} ~.
\end{equation}
The sum over $k$ is over the set indexing a choice of representatives $\{ U_k \}$ of the isomorphism classes of simple objects in $\Cc$. Unless $\Cc$ is equivalent to the category of finite-dimensional complex vector spaces, we have $R(\one) \ncong \one$, and so $R$ is not a monoidal functor. It is however still a lax and co-lax monoidal functor \cite[Prop.\,2.22]{Kong:2008ci}, and if $A$ is a symmetric Frobenius algebra in $\Cc$, then $R(A)$ is a symmetric Frobenius algebra in $\Cc_+ \boxtimes \Cc_-$ \cite[Prop.\,2.24]{Kong:2008ci}.

Given a Frobenius algebra $A$ in a modular category $\Cc$, we define the morphism $P^l_A : A \rightarrow A$ as follows \cite[Sect.\,2.4]{Frohlich:2003hm}
\begin{equation}
  P^l_A = 
  \big[ d_A \otimes \id_A \big]
  \circ \big[ id_{A^\vee} \otimes ( c_{A,A} \circ \Delta \circ m ) \big]
  \circ \big[ \tilde b_A \otimes \id_A \big] ~.
\end{equation}
Suppose $A$ is in addition symmetric and special, with $m \circ \Delta = \beta_A \id_A$. Then $\beta_A^{-1} P^l_A$ is an idempotent, and its image is a commutative subalgebra of $A$, called the left centre (see \cite[Sect.\,2.4]{Frohlich:2003hm} for details). Given a symmetric special Frobenius algebra $A$ in $\Cc$ we define the {\em full centre} $Z(A) \in \Cc_+ \boxtimes \Cc_-$ as \cite[Def.\,4.9]{unique} (see also \cite[Def.\,3.17]{Kong:2008ci})
\begin{equation}\eqlabel{ZA-def}
  Z(A) = \mathrm{im} \, P^l_{R(A)} ~.
\end{equation}
By definition, the full centre is a subobject of $R(A)$. One can restrict the algebra structure from $R(A)$ to $Z(A)$ and one finds that $Z(A)$ is a commutative symmetric Frobenius algebra \cite[App.\,A]{Runkel:2005qw}. For example, $Z(\one) = \bigoplus_k U_k^\vee \times U_k$.

If $A$ and $B$ are two symmetric Frobenius algebras in a monoidal category $\Cc$, and $f : A\rightarrow B$ is a morphism (not necessarily an algebra map), we define a morphism\footnote{
  Here we use a slightly different star symbol as for a *-category and hope this will suffice to avoid confusion.} 
$f^\star : B \rightarrow A$ as
\begin{equation}
  f^\star = 
  \big[ (\eps_B \circ m_B) \otimes \id_A \big] 
\circ \big[ \id_B \otimes f \otimes \id_A \big] 
\circ \big[ \id_B \otimes (\Delta_A \circ \eta_A)\big]~.
\end{equation}
Some properties of the map $(\,\cdot\,)^\star$ are collected in \cite[Lem.\,2.17]{Kong:2008ci}. We can now state the following definition \cite[Def.\,6.13]{Kong2006c} (the formulation used here is \cite[Def.\,3.7]{Kong:2008ci}, which also gives a graphical representation).

\begin{definition}\delabel{cardy-alg}
Let $\Cc$ be a modular category.
A {\em Cardy algebra} is a triple $(A|B,\iota)$ where $A$ is a symmetric Frobenius algebra in $\Cc$, $B$ is a commutative symmetric Frobenius algebra in $\Cc_+ \boxtimes \Cc_-$, and $\iota$ is an algebra homomorphism from $B$ to $R(A)$ such that
\\[.3em]
(i) $m_{R(A)} \circ c_{R(A),R(A)} \circ (\iota \otimes \id_{R(A)}) = m_{R(A)} \circ (\iota \otimes \id_{R(A)})$ (centre condition),
\\[.3em]
(ii) $\iota \circ \iota^\star = P^l_{R(A)}$ (Cardy condition),
\\[.3em]
(iii) $B$ is modular invariant.
\end{definition}

Here $m_{R(A)}$ denotes the multiplication on the symmetric Frobenius algebra $R(A)$. We call two Cardy algebras $(A,B,\iota)$ and $(A',B',\iota')$ {\em isomorphic} if there is a pair of Frobenius algebra isomorphisms $f: A \rightarrow A'$ and $g : B \rightarrow B'$ such that $R(f) \circ \iota = \iota' \circ g$. The following theorem states some properties of Cardy algebras. It combines \cite[Thm.\,4.26]{unique}, \cite[Thm.\,1.1]{Kong:2007yv} and \cite[Thms.\,3.18,\,3.21,\,3.22,\,3.24]{Kong:2008ci}.

\begin{theorem}\thlabel{cardy-alg}
Let $\Cc$ be a modular category.
\\[.3em]
(i) If $A$ is a special symmetric Frobenius algebra in $\Cc$ then $(A|Z(A),e)$, with $e : Z(A) \hookrightarrow R(A)$ the subobject embedding, is a Cardy algebra.
\\[.3em]
(ii) Let $(A|B,\iota)$ be a Cardy algebra such that $\dim(A) \neq 0$ and $B$ is haploid. Then $A$ is simple and special and $(A,B,\iota) \cong (A,Z(A),e)$ as Cardy algebras.
\\[.3em]
(iii) Let $B$ be a haploid modular invariant commutative symmetric Frobenius algebra in $\Cc_+ \boxtimes \Cc_-$. Then there exists a simple special symmetric Frobenius algebra $A$ in $\Cc$ and a morphism $\iota : B \rightarrow R(A)$ such that $(A|B,\iota)$ is a Cardy algebra.
\\[.3em]
(iv) Let $(A_1|B_1,\iota_1)$ and $(A_2|B_2,\iota_2)$ be two Cardy algebras such that $B_1$, $B_2$ are haploid and $\dim(A_1)$, $\dim(A_2)$ are non-zero. Then $B_1 \cong B_2$ as algebras if and only if $A_1$ and $A_2$ are Morita equivalent.
\end{theorem}

Two algebras $C$ and $D$ are Morita equivalent if there exist bimodules ${}_CM_D$ and ${}_DM_C$ such that ${}_CM_D \otimes_{\!D}\, {}_DM_C \cong C$ and ${}_DM_C \otimes_{\!C}\, {}_CM_D \cong D$.
Points (i) and (ii) imply that a symmetric special Frobenius algebra $A$ in $\Cc$ determines a Cardy algebra, and this Cardy algebra is the unique one (up to isomorphism) of the form $(A|B,\iota)$ with $B$ haploid.

\subsection{Euclidean CFT}\selabel{ecft-oc-0}

In an open/closed CFT the source categories of the functors in \deref{ecft} and \deref{gen0} are replaced by categories which contain surfaces with unparametrised boundaries. This is easiest to implement by equipping each Riemann surface with an anti-holomorphic involution and an orientation of the quotient surface. Let us describe this in more detail for $\ST$.

The objects of $\STR$ are triples $\underline{m} = (\{ 1, \dots , m \},\pi,\sigma)$, where $\pi$ is a permutation of $\{ 1, \dots , m \}$ of order two and $\sigma$ is a section of the projection onto orbits $p : \{ 1, \dots , m \} \rightarrow \{ 1, \dots , m \} / \langle \pi \rangle$, i.e.\ $p \circ \sigma = \id$. 
A morphism in $\Hom_{\STR}(\underline{m},\underline{n})$ consists of a morphism $\hat\Sigma \in \Hom_{\ST}(m,n)$, together with an anti-holomorphic involution $\iota : \hat\Sigma \rightarrow \hat\Sigma$ and an orientation of the quotient surface $\Sigma = \hat\Sigma / \langle \iota \rangle$. The involution $\iota$ has to be compatible with the punctures, namely, if $(p,U,\varphi)$ is a puncture, so is $(\iota(p),\iota(U),C \circ \varphi \circ \iota)$, where $C(z) = z^*$ is complex conjugation. Thus $\iota$ induces a permutation of order two on the sets of punctures. This permutation has to agree with those in $\underline{m}$ and $\underline{n}$. By fixing an orientation on the quotient $\Sigma$ we in particular insist that it be orientable. Also, since $\Sigma$ carries a conformal structure coming from $\hat\Sigma$, adding an orientation makes $\Sigma$ into a Riemann surface with boundary. 
Denote the projection $\hat\Sigma \rightarrow \Sigma$ by $\tau$. The orientation of $\Sigma$ defines an open subset $\Sigma^+ \subset \hat\Sigma$ consisting of all points such that $\tau$ is orientation preserving in a neighbourhood of that point. If for a puncture $(p,U,\varphi)$ we have $\iota(p)=p$, then we demand that $\varphi$ maps $U \cap \Sigma^+$ to the upper half plane. Finally, if a puncture in $\Sigma^+$ is assigned an object label $k \in \{1,\dots,m\}$ (or $\{1,\dots,n\}$) then $k$ has to lie in the image of the section $\sigma$. These somewhat complicated rules ensure that the glued surface again has an involution and an orientation of the quotient.
Altogether $\STR$ is a partial symmetric monoidal category (see \cite[Sect.\,5]{osvoa} and \cite[Sect.\,3]{Kong2006c} for a description using Swiss-cheese partial (di)operads).

The `physically relevant surface', e.g.\ the world sheet of an open string or the surface obtained in the continuum limit of a lattice model, is the quotient $\Sigma$, not its double $\hat\Sigma$.

There is an embedding of $\ST$ into $\STR$ given by mapping a sphere with tubes in $\ST$ to the disjoint union of this sphere and its complex conjugate, together with the exchange of the two copies as involution. Let $\mathbb{A}_\lambda$ be the image of the annulus $A_\lambda \in \Hom_{\ST}(1,1)$ under this embedding. As another example of an object of $\STR$, the surface $A_\lambda$ can itself be equipped with an involution such that the quotient is a disc with two boundary punctures.

In more detail, for $r \in \Rb$, $r>0$, let $\mathbb{S}_r \in \Hom_{\STR}(\underline{1},\underline{1})$ ($\mathbb{S}$ for `strip') be $A_r$ with involution $\iota(z)=z^*$ and the quotient $\Cb \cup \{ \infty \} / \langle \iota \rangle$ identified with the upper half plane (with infinity and the real line); the orientation is that of the upper half plane. 
Denote by $(12) \in \STR$ the object $(\{1,2\}, \pi_{12},\sigma_1)$ with $\pi_{12}$ the transposition of two elements and $\sigma_1$ the section whose image is $\{1\}$.
For $\lambda \in \Cb^\times$, $\mathbb{A}_\lambda \in \Hom_{\STR}((12),(12))$ is given by $\mathbb{A}_\lambda = A_\lambda \sqcup A_{\lambda^*}$ with involution $\iota(z_1) = z_2^*$, $\iota(z_2) = z_1^*$, where $1$, $2$ refer to the two copies. We identify the quotient with $A_\lambda$ and take its orientation. 

\begin{definition}\delabel{oc-gen0}
An {\em open/closed oriented projective Euclidean genus-0 CFT} (genus-0 boundary CFT for short) is a smooth projective symmetric monoidal functor $\PF : \STR \rightarrow \GV$ such that
\\[.3em]
(i) $A=\PF( \underline{1} )$ is $\Rb$-graded, and $\PF(\mathbb{S}_r)$ is the line spanned by the linear map that acts on the graded component $A_{h} \subset B$ by multiplication with $r^{-h}$,
\\[.3em]
(ii) $B=\PF((12))$ is $\Rb \times \Rb$-graded, and $\PF(\mathbb{A}_\lambda)$ is the line spanned by the linear map that acts on the graded component $B_{h_l,h_r} \subset B$ by multiplication with $\lambda^{-h_l} (\lambda^*)^{-h_r}$.
\end{definition}

\begin{theorem}\thlabel{oc-gen0}
Let $V$ be a rational vertex operator algebra and let $\Cc = \mathrm{Rep}(V)$. A triple $(A|B,\iota)$ with $A \in \Cc$ and $B \in \Cc_+ \boxtimes \Cc_-$ satisfying all conditions in \deref{cardy-alg} except for (ii) and (iii) gives rise to a genus-0 boundary CFT $\PF : \STR \rightarrow \GV$ with $\PF( \underline{1} ) = A$ and $\PF((12))=B$.
\end{theorem}

This follows from \cite[Thm.\,3.17\,\&\,Sect.\,6]{Kong2006c}. Again it should be possible to formulate a converse statement, but this has so far not been done in the present setting. However, there is a converse statement in a different approach based on three-dimensional topological field theory on which we will comment in the end of this section.

\medskip

The functor $\PF$ in \thref{oc-gen0} assigns the algebra $A$ to boundary punctures and the algebra $B$ to interior punctures in the following sense. The permutation in $\underline{1} \equiv ( \{1 \} , \pi=\id, \sigma=\id)$ acts trivially, and by the compatibility condition on the permutation and the involution of a morphism in $\STR$ the element $1$ can only be assigned to a puncture that lies on a line of fixed points of the involution, i.e.\ on the boundary of the quotient surface. By the same argument, the two elements $1,2$ in $(12) \equiv ( \{1,2\} , \pi_{12}, \sigma_1 )$ are assigned to punctures that are not left fixed and hence lie in the interior of the quotient surface. 

\subsubsection*{Genus 1}

As in \seref{ecft-c-0} we can try to allow higher genus Riemann surfaces with punctures instead of just spheres, enlarging the category $\STR$ to $\RTR$. A smooth projective symmetric monoidal functor $\tilde\PF : \RTR \rightarrow \GV$ will be uniquely determined by the underlying projective functor $\PF : \STR \rightarrow \GV$, but not every such $\PF$ can be extended to $\RTR$. As in \seref{ecft-c-0} one can derive necessary conditions for this to be the case. Two such conditions are modular invariance and the Cardy condition: Modular invariance was already discussed in \seref{ecft-c-0}, and the Cardy condition \cite{Cardy:1989ir,Lewellen:1991tb} is obtained by comparing two ways to obtain an annulus with two boundary punctures, namely by sewing a disc with four boundary punctures and by sewing two discs with one boundary and one interior puncture each (the descriptions refer to the quotient surfaces). 

\begin{theorem}\thlabel{mod-cardy} \cite[Thm.\,6.15]{Kong2006c}
A genus-0 boundary CFT $\PF : \STR \rightarrow \GV$ obtained from a triple $(A|B,\iota)$ via \thref{oc-gen0} is compatible with modular invariance and the Cardy condition if and only if $(A|B,\iota)$ is a Cardy algebra.
\end{theorem}

If we choose the vertex operator algebra to be trivial, $V= \Cb$, then $\mathrm{Rep}(V)$ is the category of finite-dimensional complex vector spaces, and the above theorem describes two-dimensional oriented open/closed {\em topological} field theories over $\Cb$. Such theories have been investigated in \cite{Laz,AN,LP,MSeg}.
On the algebraic side, the main structural differences to the conformal case are that in the latter (cf.\ \deref{cardy-alg}) the algebras $A$ and $B$ live in different categories, and that the modular invariance condition (iii) appears.

\medskip

The genus 1 conditions in \deref{cardy-alg} (requirements (ii) and (iii)) turn out to be quite powerful. For example, as remarked below \thref{cardy-alg}, if in \thref{oc-gen0} $A$ is in addition special and we demand $B$ to be haploid (i.e.\ there is a unique vacuum), then the genus-0 boundary CFT is uniquely determined by $A$ alone.
Furthermore, \thref{cardy-alg}\,(iii) means that every genus-0 CFT obtained via \thref{csFa-closed-0} with $V_L = V_R = V$ such that $B$ is haploid and satisfies modular invariance actually forms part of a genus-0 boundary CFT obtained via \thref{oc-gen0}.

\subsubsection*{Relation the three-dimensional topological field theory}

One can ask if the necessary conditions stated in \thref{mod-cardy} are sufficient. As in the closed case this is not known because at the moment one cannot control the sewing of higher genus surfaces. However, the arguments in \cite{Lewellen:1991tb} and the approach in \cite{Felder:1999cv,tft0,tft1,tft5} via three-dimensional topological field theory (3d\,TFT) support that there will be no additional conditions on the algebraic side. While the arguments in \cite{Lewellen:1991tb} and in \thref{mod-cardy} are based on generators and relations, the 3d\,TFT provides an `a priori' construction, which we briefly outline.

In the 3d\,TFT approach the chiral 2d\,CFT lives on the boundary $\hat X = \partial M$ of a three-manifold $M$ with embedded ribbon graph. 
The relevant 3d\,TFT is the one constructed from the modular category $\Cc = \mathrm{Rep}(V)$, for $V$ the rational vertex operator algebra describing the chiral subtheories.
The 3d\,TFT assigns to the two-dimensional boundary $\hat X = \partial M$ a vector space $H(\hat X)$, interpreted as the space of conformal blocks\footnote{
  Roughly speaking, for a given $\hat\Sigma \in \Hom_{\STR}(\underline{m},\emptyset)$ the conformal blocks for $\hat  \Sigma$ are all linear maps from $\PF(\underline{m}) \in \GV$ to $\PF(\emptyset) = \Cb$ compatible with the chiral symmetry described by the vertex operator algebra $V$, cf.\ \cite[Ch.\,9\,\&\,10]{Frenkel-BenZvi}.
}, and to the three-manifold $M$ a vector $C_{\hat X} \in H(\hat X)$, interpreted as $\PF(\hat X)$. This defines all $\PF(\hat X)$ simultaneously. One then proves that this assignment is compatible with sewing \cite[Sect.\,2]{tft5}. The construction requires the choice of a special symmetric Frobenius algebra $A$ in the modular category that defines the 3d\,TFT. Indeed, this is the context in which the relevance of such Frobenius algebras to boundary CFT first became apparent \cite{tft0}. In the language used here, the 2d\,CFT constructed by the 3d\,TFT approach is the one defined by the Cardy algebra $(A|Z(A),e)$, cf.\ Theorems \ref{th:cardy-alg}\,(i) and \ref{th:oc-gen0}.

Conversely, in \cite[Sect.\,3.4]{unique} the constraints on a collection of vectors $C_{\hat X} \in H(\hat X)$ (one for each $\hat X$) to be compatible with sewing are formulated as a monoidal natural transformation between two functors, and it is proved in \cite[Thm.\,4.26]{unique} that under certain natural conditions (listed there), each solution to the sewing constraints is equivalent to one obtained from a special symmetric Frobenius algebra $A$ in $\Cc$.

\subsection{Minkowskian CFT}\selabel{mcft-oc}

A boundary CFT in Minkowskian spacetime is described by a net of operator algebras on Minkowski half-space $M_+ = \{ (x,t) \in M_2 \,|\, x>0 \}$. A double cone in $M_+$ is a double cone in $M_2$ whose closure is contained in $M_+$. In other words, $O = I \times J$ is a double cone in $M_+$ iff $J = (a,b)$ and $I=(c,d)$ with $a<b<c<d$. The conformal group acting on $M_2$ is reduced to those transformation which preserve $M_+$. This is the diagonal $\mathrm{PSL}(2,\Rb)$, namely, $g \in \mathrm{PSL}(2,\Rb)$ acts on a double cone as $gO = gI \times gJ$.
When referring to nets on $S^1$ below, it is understood that the $S^1$ is identified with the one-point compactification of the time axis $\{ x=0 \}$ via $\zeta \mapsto i(1-\zeta)/(1+\zeta)$.

\begin{definition} \delabel{m-bCFT} \cite[Def.\,2.2]{Longo:2004fu}
Let $\Vc$ be an irreducible M\"obius covariant net on $S^1$. A {\em Min\-kowskian boundary CFT associated with} $\Vc$ is a tuple $(\Hc,\Ac_+,\Omega,U,\pi)$, where $\Hc$ is a separable Hilbert space, $\Ac_+ : O \mapsto \Ac_+(O)$ assigns von Neumann algebras in $B(\Hc)$ to double cones in $M_+$, $U$ is a strongly continuous projective unitary positive energy representation of $\mathrm{PSL}(2,\Rb)$ on $\Hc$, $\Omega \in \Hc$ is invariant under that action, and $\pi$ is a representation of $\Vc$ on $\Hc$ such that $\pi(\Vc(I) \vee \Vc(J)) \subset \Ac_+(I \times J)$. These data obey the properties (i)--(v) in \deref{mcft} (with $\Bc$ and $\mathrm{PSL}(2,\Rb)\times\mathrm{PSL}(2,\Rb)$ replaced by $\Ac_+$ and  $\mathrm{PSL}(2,\Rb)$)\footnote{
  In \cite{Longo:2004fu} the authors are interested in the case that $\Vc$ is completely rational. 
  Additivity (property (iii) in \deref{mcft}) is then a consequence (cf.\ \cite[Prop.\,2.12]{Longo:2004fu}) 
  and is not included in \cite[Def.\,2.2]{Longo:2004fu}.
}. In addition we have:
\\[.3em]
(i) Every vector invariant under the action of $\mathrm{PSL}(2,\Rb)$ is proportional to $\Omega$.
\\[.3em]
(ii) $\pi$ is covariant, i.e.\ $U(g) \pi(\Vc(I)) U(g)^* = \pi(\Vc(gI))$ for all $g\in\mathrm{PSL}(2,\Rb)$ and $I$ an interval in $\Rb$ such that also $gI \subset \Rb$.
\\[.3em]
(iii) Let $\pi(\Vc_+)$ be the $C^*$-algebra generated by $\pi(\Vc(I) \vee \Vc(J))$ for all double cones $O = I \times J$ in $M_+$. Then for each double cone $O$, $\Ac_+(O) \vee \pi(\Vc_+)''=B(\Hc)$. 
\end{definition}

A boundary CFT $\Ac_+$ is called {\em Haag dual} if for all double cones $O \in M_+$ one has $\Ac_+(O) = \Ac_+(O')'$. Here $\Ac_+(O') = \bigvee_{\!Q}\, \Ac_+(Q)$ where $Q$ runs over all double cones in $M_+$ that are space-like separated from $O$.

\begin{theorem}\thlabel{m-bCFT}
Let $\Vc$ be a completely rational net on $S^1$. Haag dual Minkowskian boundary CFTs $\Ac_+$ associated with $\Vc$ are in one-to-one correspondence to haploid\footnote{
  From the Euclidean point of view `simple' would be the more natural condition here, as it is preserved by Morita equivalence. There, every algebra in a given Morita class can serve as a boundary theory for a given closed CFT \cite[Sect.\,4.1]{tft1}. The occurrence of haploidity in \thref{m-bCFT} is again tied to the uniqueness of the vacuum, condition (i) in \deref{m-bCFT}.
  } 
special symmetric *-Frobenius algebras $A \in \mathrm{Rep}(\Vc)$.
\end{theorem}

The theorem follows from \cite[Thm.\,4.9]{Longo:1994xe} and \cite[Prop.\,2.9]{Longo:2004fu}. It is also shown in \cite[Sect.\,2.3]{Longo:2004fu} that all boundary CFTs associated with $\Vc$ are subtheories of a Haag dual boundary CFT. 

A boundary CFT $\Ac_+$ can be used to define a net of operator algebras on $\Rb$ by assigning to an interval $K \subset \Rb$ the von Neumann algebra $\Ac(K)$ generated by all $\Ac_+(I \times J)$ with $I,J \subset K$. The algebra $A$ in \thref{m-bCFT} describes the chiral extension $\Ac$ of the net $\Vc$ via \cite[Thm.\,4.9]{Longo:1994xe}. Conversely, if one is given $\Ac$ one can define a net on $M_+$ by setting $\Ac_+(I \times J) = \Ac(L) \cap \Ac(K)'$, where, for $a<b<c<d$, $J=(a,b)$, $I=(c,d)$, $L=(a,d)$ and $K=(b,c)$ (see \cite[Fig.\,5]{Longo:2004fu}). This net is automatically Haag dual \cite[Prop.\,2.9]{Longo:2004fu}.

From the data in \thref{m-bCFT} one can also construct a Minkowskian CFT on $M_2$ using the methods in \cite{Bockenhauer:1999ae,Rehren:1999rj}.

\begin{theorem}\thlabel{m-ssFA-CFT} \cite[Cor.\,1.6]{Rehren:1999rj}
Let $\Vc$ be a completely rational net on $S^1$. A haploid special symmetric *-Frobenius algebras $A \in \mathrm{Rep}(\Vc)$ gives rise to a Minkowskian CFT $\mathcal{B}$ with subtheory $\Vc \otimes \Vc$.
\end{theorem}

The CFTs $\Ac_+$ on $M_+$ and $\Bc$ on $M_2$ defined by the same Frobenius algebra $A$ via Theorems \ref{th:m-bCFT} and \ref{th:m-ssFA-CFT} are locally isomorphic in the sense explained in \cite[Thm.\,4.1]{Longo:2004fu}.

\medskip

By \thref{m-c-frob}, the CFT $\Bc$ in turn corresponds to a haploid commutative symmetric *-Frobenius algebras $B_A$ in $\mathrm{Rep}(\Vc)_+ \boxtimes \mathrm{Rep}(\Vc)_-$, determined by $A$. 
Comparing \cite[Cor.\,1.6]{Rehren:1999rj} and \cite[Eqn.\,(3.9)]{Runkel:2005qw} we see that $B_A \cong Z(A)$ as objects. To establish that the algebra structure agrees one has to compare the construction in \cite[Sect.\,3]{Rehren:1999rj} to the definition of $Z(A)$ in \equref{ZA-def}. We did not carry out this comparison in detail, but (not only) we are convinced that indeed $B_A  \cong Z(A)$ as algebras.

With this caveat, we can note that as in the Euclidean setting, a special symmetric Frobenius algebra $A$ determines a boundary CFT, such that $A$ is associated to the boundary punctures/observables, and $Z(A)$ to the interior punctures/observables. 

We can also ask if a Min\-kowskian CFT $\Bc$ on $M_2$ occurs as the observables of a Minkowskian boundary CFT in the sense of \thref{m-ssFA-CFT}. If $\Bc$ contains $\Vc \otimes \Vc$ as a subtheory, for a completely rational net $\Vc$, the answer is provided by \thref{cardy-alg}: Let $\Cc = \mathrm{Rep}(\Vc)$. $\Bc$ is described by a haploid commutative symmetric *-Frobenius algebra $B \in \Cc_+ \boxtimes \Cc_-$. There exists a special symmetric Frobenius algebra $A \in \Cc$ such that $B \cong Z(A)$ as algebras if and only if $\dim(B) = \mathrm{Dim}(\Cc)$ (this in turn is true if and only if $B$ is modular invariant). If $A$ exists, it is simple (and can be chosen haploid \cite[Prop.\,4.10]{Kong:2007yv}) and provides\footnote{
  One needs to check that $A$ is a *-Frobenius algebra. Looking at the proof of \cite[Thm.\,3.22]{Kong:2008ci} we see that $A$ is a direct summand of $T(B)$, and the functor $T$ preserves the *-structure. Since $B$ is a *-Frobenius algebra, so is $A$.
  }
a boundary CFT with $\Bc$ as associated CFT on $M_2$ via Theorems \ref{th:m-bCFT} and \ref{th:m-ssFA-CFT}.


\subsubsection*{Genus 1}

The physical interpretation of the Cardy condition in Minkowski spacetime seems even more surprising (and less natural to demand) than modular invariance. The Cardy condition (without boundary insertions) amounts to comparing a quantum system on an interval of length $R$ in a Gibbs state of inverse temperature $\beta$ to the same system on a circle of radius $\beta$ in a `thermal spectrum' of particles, cf.\ \cite{Ghoshal:1993tm,Hannabuss:2003va}. 

The point made in \cite[App.\,C]{Longo:2007zz} is that modular invariance and the Cardy condition hold automatically if (and only if) the boundary CFT is Haag dual. This is a natural maximality condition for a quantum field theory, as it states that the observables localised in the region which is spacelike separated from a double cone $O$ are precisely those operators that commute with all observables localised in $O$.

\section{Conclusion}\selabel{conc}

We have pointed out some places where the Euclidean and Minkowskian approach to two-dimensional conformal field theory lead to the same algebraic structures. 

One lesson to learn from this is that it is worth stating and proving theorems in the algebraic setting of monoidal categories without making reference to an underlying realisation via representations of a rational vertex operator algebra or a completely rational conformal net. In this way, results apply in the Euclidean and Minkowskian setting alike. The existence of an open/closed CFT for a given closed CFT with the same rational left and right chiral symmetry as implied by \thref{cardy-alg} is an example of this.

Of course the link of Euclidean and Minkowskian CFT via modular categories is somewhat indirect, and it would be desirable to have theorems relating the two approaches directly.

\bigskip\noindent
{\bf Acknowledgements:} IR is grateful to the organisers of the stimulating conferences
``Non-commutative Structures in Mathematics and Physics'', Brussels (July 22--26, 2008), and
``Operator Algebras, Conformal Field Theory and Related Topics'', Vienna (September 8--19, 2008), for inviting him and giving him the opportunity to speak.
We thank J\"urgen Fuchs, Yi-Zhi Huang, Karl-Henning Rehren and Christoph Schweigert 
for helpful comments on a draft version of these proceedings.
LK is supported in part by the Gordon and Betty Moore
Foundation through Caltech's Center for the Physics of Information, and
by the National Science Foundation under Grant No. PHY-0803371.
IR is in part supported by the EPSRC First Grant EP/E005047/1 and the STFC Rolling Grant ST/G000395/1.

\newcommand\arxiv[2]      {\href{http://arXiv.org/abs/#1}{#2}}
\newcommand\doi[2]        {\href{http://dx.doi.org/#1}{#2}}
\newcommand\httpurl[2]    {\href{http://#1}{#2}}


\end{document}